# Chaotic dynamics of the Hunt model, an artificially constructed flow system with a hyperbolic attractor

*Yu.S. Aidarova, S.P. Kuznetsov*
Kotel'nikov's Institute of Radio-Engineering and Electronics, Saratov Branch
Saratov State University

We study numerically chaotic behavior associated with a hyperbolic strange attractor of Plykin type in the model of Hunt, an artificially constructed dynamical system with continuous time. There are presented portraits of the attractor, plots of realizations for chaotic signal generated by the system, illustrations of the sensitive dependence on initial conditions for the trajectories on the attractor. Quantitative characteristics of the attractor are estimated, including the Lyapunov exponents and the attractor dimension. We discuss symbolic dynamics on the attractor, find out and analyze some unstable periodic orbit belonging to the attractor.

## 1.Introduction

In mathematical theory of dynamical systems a class of uniformly hyperbolic strange attractors is introduced [1-9]. In such an attractor all orbits are of saddle type, and their stable and unstable manifolds do not touch each other, but can only intersect transversally. These attractors manifest strong stochastic properties and allow detailed mathematical analysis. They are structurally stable, that is insensitive in respect to variation of functions and parameters in the dynamical equations. The main concepts or the mathematical theory were developed more than 40 years ago, but until recent times, the uniformly hyperbolic strange attractors were regarded only as purified images of chaos, not intrinsic for realistic models of systems with complicated dynamics.

In textbooks and reviews, examples of the uniformly hyperbolic attractors are traditionally represented by mathematical constructions, the Plykin attractor and the Smale – Williams solenoid. These examples relate to discrete-time systems, the iterated maps. The Smale – Williams attractor appears in a map of a toroidal domain into itself in phase space of dimension 3 or more. The Plykin attractor appears in some special mapping of a bounded domain on a plane with three holes.

In applications, physics and technology people deal more often with systems operating in continuous time called the flows in mathematical literature. The passage from a map $\mathbf{x}_{n+1} = \mathbf{f}(\mathbf{x}_n)$ to a flow system is called the *suspension* [2-7]. Such a passage is possible if the map is invertible. For the resulting flow system the relation $\mathbf{x}_{n+1} = \mathbf{f}(\mathbf{x}_n)$ is called the Poincaré map, or the stroboscopic map in the context of non-autonomous systems [10-12].

Recently, a system was suggested and realized experimentally, in which the Poincaré map possesses the attractor of Smale – Williams type [13, 14]. It is composed of two non-autonomous van der Pol oscillators, which become active turn by turn and transfer the excitation each other, and transformation of the phase of oscillations on a whole cycle corresponds to the expanding circle map. Results of computer verification of conditions of the theorem guaranteeing existence of the hyperbolic attractor were presented in Refs. [15,16]. Other variants of analogous schemas based on autonomous and non-autonomous oscillators are discussed in Refs. [17-19].

For attractors of Plykin type no physical examples were introduced, although in Ref. [20] the authors argue in favor of existence of such an attractor in the Poincaré map of an autonomous three-dimensional set of differential equations relating to the neuron model. On the other hand, an explicit example of a non-autonomous flow system with Plykin type attractor in the stroboscopic map was advanced in the PhD thesis of T. Hunt under supervision of prof. Robert MacKay in Cambridge university [21]. The model of Hunt is defined by means of multiple expressions distinct for different domains in the state space and contains many artificially introduced smoothing factors. (The flow relates to the class C$^1$, that means that the solutions are continuous together with the first derivatives.) It is really hard to imagine that this model could be implemented on a base of some physical system.

Nevertheless, elaboration of the Hunt model, as a continuous time system, may be regarded as a productive step towards construction of real examples with hyperbolic strange attractors.

In this work, we intend to reproduce the Hunt construction and undertake numerical studies of the dynamics of the system exploiting tool-box of well-elaborated methods of nonlinear dynamics and techniques of presentation, including phase portraits, time dependences for variables, computation of Lyapunov exponents and dimensions. One of the main goals is to develop methodology and get experience of dealing with hyperbolic attractors, including application of numerical procedures for verifying hyperbolicity, based on the so-called cone criterion.

## *2. Qualitative description of the Hunt model*

Hunt model is a non-autonomous system governed by differential equations for two variables $x$ and $y$, with right-hand parts depending on $x$, $y$ and time variable $t$:

$$dx/dt = f_*(x,y,t), \quad dy/dt = g_*(x,y,t). \qquad (1)$$

Here the functions $f_*$ and $g_*$ are continuous, differentiable, and $2\pi$-periodic in respect to the argument $t$. Formal description of the Hunt model and mathematical relations for computation of the functions $f_*$ and $g_*$ are given in Appendix A.

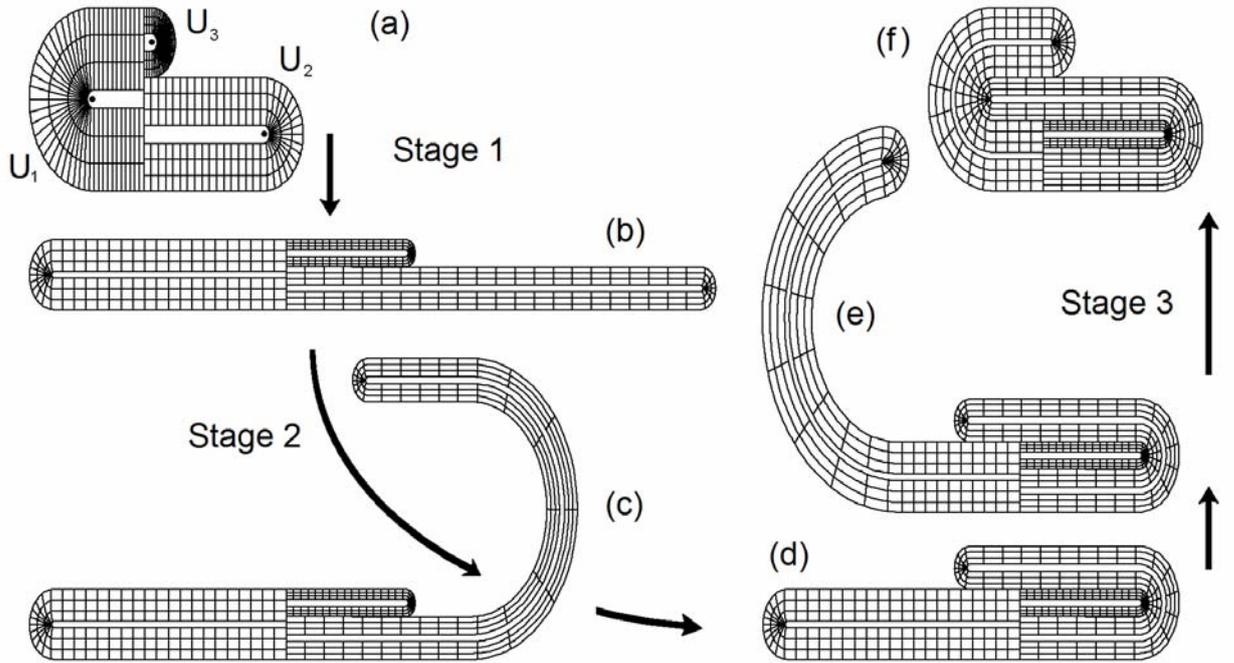

**Fig.1.** Evolution of the domain $U = U_1 \cup U_2 \cup U_3$ on the plane $x$, $y$ on one period $\Delta t = 2\pi$ in the Hunt model. The filled small black circles on the panel (a) indicate origins of the special curvilinear coordinates

Figure 1a shows the initial (at $t = 0$) configuration of the domain, dynamics of which is monitored in the construction under consideration. This area $U$ contains three sub-areas $U_i$, in which we define, following Hunt, special curvilinear coordinates, $(r,\theta)^i$, $i$=1,2,3. The respective families of coordinate lines are shown in the figure. In the course of evolution in time, the points associated with the origins move in a definite way, while the domains themselves and the coordinate curves being continuously deformed.

The dynamics of variables $x,y$ in the time interval $2\pi$ is represented as three consecutive stages, the duration of each is $2\pi/3$. At the first stage the area $U$ undergoes horizontal expansion and vertical compression, as shown in Fig.1b. (More accurately, compressing and stretching take place along the coordinate lines of the curvilinear coordinate system, $\theta = \text{const}$ and $r = \text{const}$, respectively.) In the second stage, the sub-area $U_2$ disposed at the right side of the figure smoothly bends up, left and down (Fig. 1c), so that becomes placed along the border of the domain $U_3$ (Fig. 1d). At the third stage the sub-area $U_1$ in the left part of the figure undergoes similar deformations, bending up, right, and down (Fig.1e). The result is that its lower border is placed on the edge of the domain $U_2$ transformed in the previous stage (Fig.1f). The whole area takes the form of the original domain. Moreover, the coordinate lines in their new position run along the lines of the original coordinates. The deformation described represents, to say, the main content of dynamics. Besides, the model contains some additional modifications, by which the points of the origin of the curvilinear coordinates become repelling, and transverse compression of the figure is stronger, with the result that the transformed domain appears to be inside the original area (with match of the lower boundaries for the original region and its image). The degree of smoothing is determined by a parameter whose value in the following computations will be taken $\varepsilon = 0.17$. Also, the model provides a definition for the vector field associated with the right parts of the equations (1) outside the deforming domain $U(t)$.

## 3. Attractor of the Hunt model and its properties

Integrating Eqs. (2) on a time interval $\Delta t = 2\pi$ with initial conditions $\mathbf{x}_0 = (x_0, y_0)$ we get a new state vector $\mathbf{x}_1 = (x_1, y_1)$ that may be treated as a result of application of some map $\mathbf{T}$ to the original vector $\mathbf{x}_0$. This is the Poincaré map, or the stroboscopic map for the flow under consideration. With initial vector in the domain U the image is also in U. In other words, the image $\mathbf{T}(U)$ is a subset of U, and attractor of the map $\mathbf{T}$ may be defined as intersection of the image sets obtained from multiple application of the map: $A = \bigcap_{k=1}^{\infty} \mathbf{T}^k(U)$. For the discrete-time system associated with the map $\mathbf{T}$, this is a hyperbolic attractor of Plykin type [1-7,22]. In the context of the model under consideration, the mathematical foundation of the hyperbolicity of the attractor is presented by Hunt. In Appendix B we consider an alternative approach to verification of this property by means of computations.

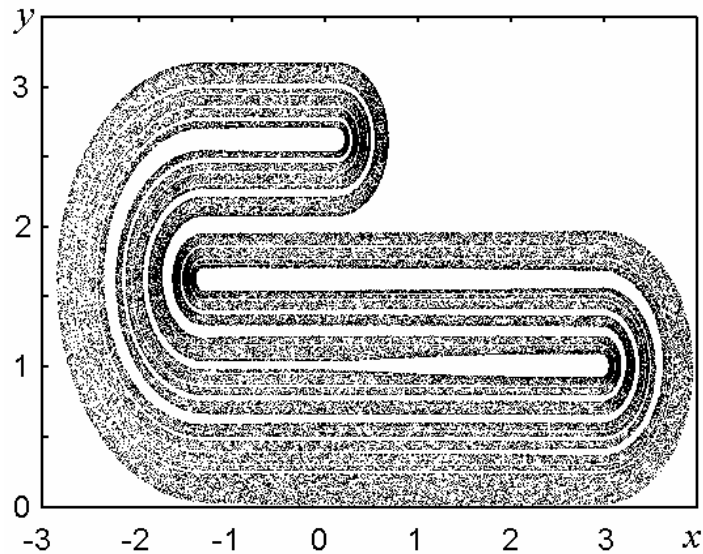

**Fig. 2.** Attractor of the Hunt model in the Poincare cross-section

For numerical solution of the equations we used the Runge – Kutta method of the 4-th order. As the functions in the definition of the model do not have a high degree of smoothness, the method inevitably loses accuracy. Nevertheless, as found empirically, the results with this method are better than those with methods of lower order. To get higher accuracy, we simply decrease the integration step.

Figure 2 shows the portrait of the attractor in the Poincare cross-section obtained for the Hunt model from numerical solution of Eqs. (1). It is a set of points of a representative trajectory on the attractor at the instances of time being multiple $2\pi$, in large enough number (some tens thousand). The attractor is characterised by presence of transversal fractal structure. Some first levels of this structure are well visible: the object consists of strips, each of which contains strips of a next level, and so far, *ad infinitum*.

Since the main reason for the development of Hunt's model is to build a system with continuous time with a hyperbolic attractor, it is natural to turn to the illustrations of the distinctive features of the flow system.

Figure 3 depicts attractor of the flow system in projection on the plane ($x$, $y$). It is remarkable that it is composed of two "butterfly wings", like the Lorenz attractor, although they are asymmetrical. Formation of these "wings" corresponds to the second and the third stage of the construction in the Hunt model, which are accompanied with deformations and rotations of the side sub-areas (see Fig. 1). Figure 4 plots are shown for variables $x$ and $y$ versus time. Observe erratic behavior on sufficiently large times, indicating chaotic nature of the dynamics. The dependences are continuous, but on each period of duration $2\pi$ they are characterized by presence of matched intervals, the horizontal plateau and sudden bursts. This is specific for the Hunt model and reflects the dynamics built of successive stages.

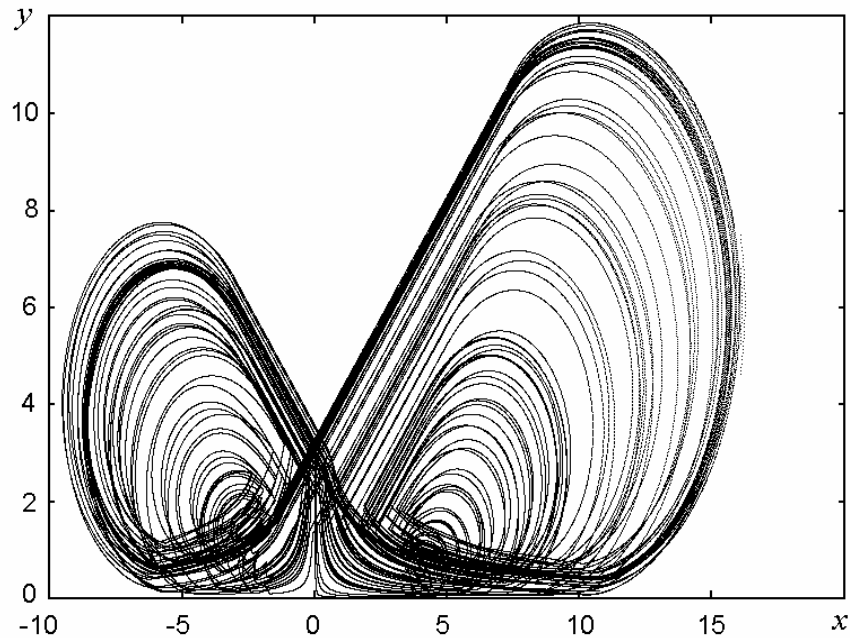

**Fig. 3.** Attractor of Hunt model in projection on the plane of variables $x$ and $y$.

Figure 5 shows the power spectrum for the variable $x$. On the vertical axis the logarithmic scale is used. According to methodology usual in the applied spectral analysis of random processes [23], the existing sample of the process $x(t)$ is subdivided in sections of length $T_0$, each segment is multiplied by $\sin^2 \pi\tau/T_0$ (the so-called "window"), then the Fourier transformation is applied, and the result is averaged over all the sections of the signal. The value $\Delta\omega \cong \pi/T_0$ determines the resolution of the spectral analysis. The greater the number of the sections of the signal, the smaller the mean error of estimate for the spectral power density.

The signal was represented by time series with step of sampling $\Delta t = 2\pi/120$, each section $T_0$ containing $2^{13}$ samples, and the averaging was performed over 64 sections of the signal. As seen from the diagram, the spectrum looks continuous, i.e. it is of the same nature as spectra of stationary random processes. Some peaks at frequencies 1, 2, 3 ... arise because of periodicity intrinsic to the evolution rule in the Hunt model.

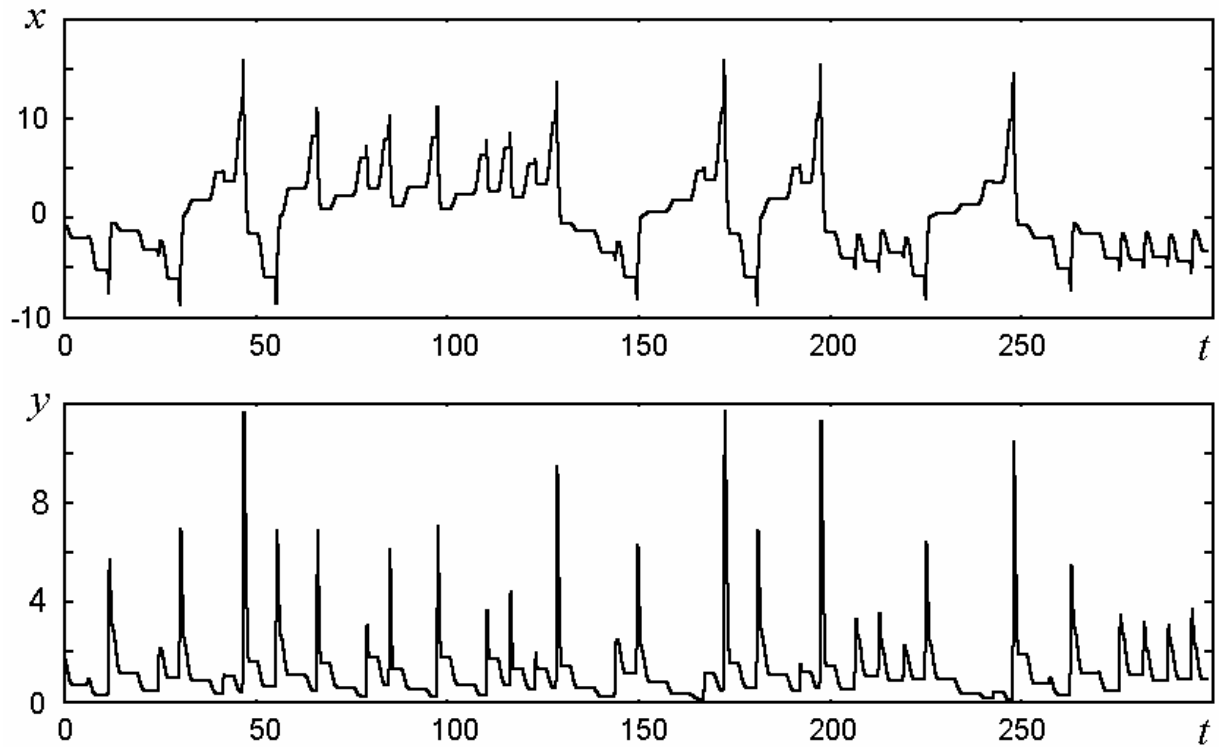

**Fig. 4.** Dynamical variables *x* and *y* versus time in the Hunt model

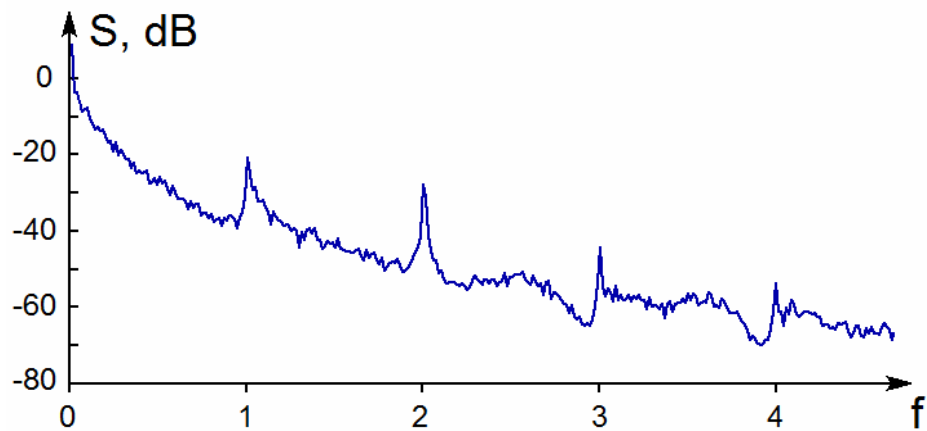

**Fig. 5.** Spectral power density for a signal generated by the Hunt model on the attractor.

As known, one of the key attributes of the dynamical chaos is high sensitivity of the phase trajectories to small variations of initial conditions. To demonstrate this property in the Hunt model, we perform multiple numerical integration of equations (1) setting initial conditions each time at a certain point on the attractor with a small random perturbation of *x* and *y*. The results are presented as overlay of 20 samples of the process in one plot, see Fig. 6. Observe that initial parts of the samples reproduce each other with rather high accuracy being visually indistinguishable, but over time, closer to the right side of the diagram, they differ from each other all the stronger, and the picture smears out. Periodic bursts on this background persist due to the mentioned specifics of the Hunt model, the non-autonomous dynamical system with time-periodic right-hand parts of the equations.

For the quantitative characteristics of instability of trajectories inherent to chaos the Lyapunov exponents are used. In our case, there are two non-trivial exponents, one positive, responsible for the instability of the motion on the attractor, and the other negative, corresponding to the approach of the trajectories to the attractor.

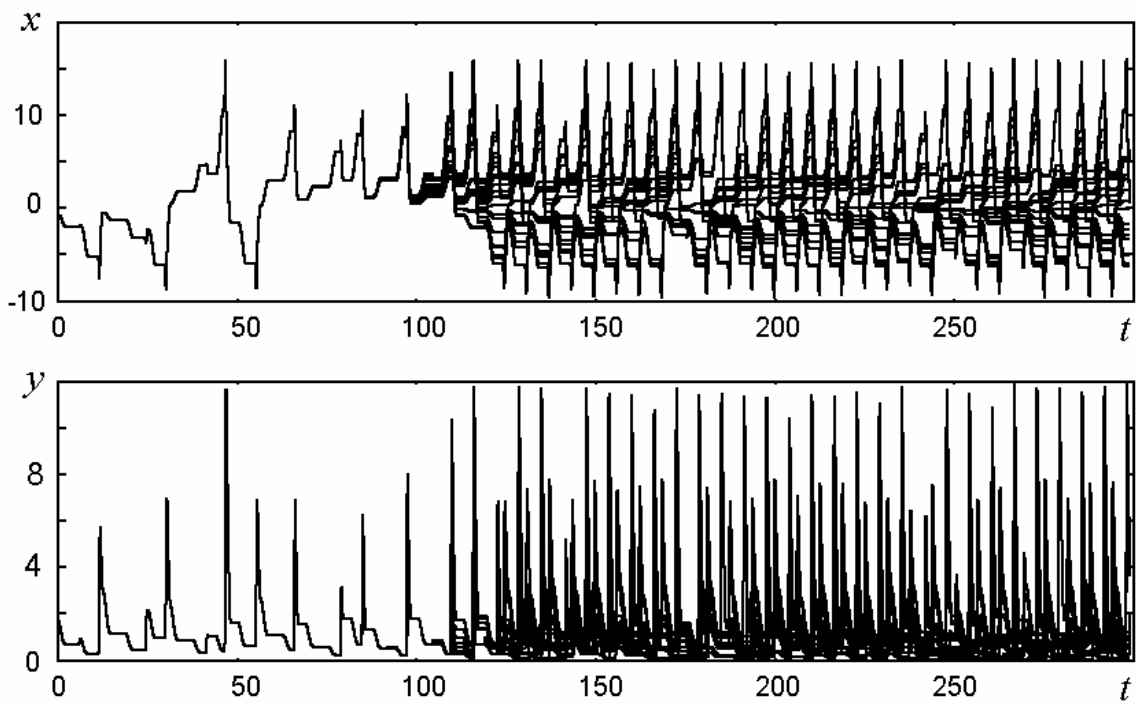

**Fig. 6.** Illustration of instability of the orbits on the Hunt attractor in respect to small variation of initial conditions. A set of 20 samples for the dependencies of the variables on time are superimposed with slightly different initial conditions.

For computation of the Lyapunov exponents, the Benettin algorithm was used [24,12]. Because of complexity of the formal definition of the Hunt model, we prefer a version of the algorithm without derivation of equations in variations. Namely, we integrate collection of three sets of Eqs. (1). One corresponds to motion of a representative point on a main phase trajectory, and two others to close neighboring orbits, with fixed initial norm of the perturbation vector $\sqrt{(x_i - x_0)^2 + (y_i - y_0)^2} = \varepsilon_0 \ll 1$, $i$=1,2. After each time interval of duration $2\pi$ the program performs orthogonalization of the vectors with the Gram – Schmidt process and normalizes the perturbation vectors to get again the norms $\varepsilon_0$. The Lyapunov exponents are estimated as average rates for growth or decrease of accumulated sums of logarithms for norm ratios at the end and at the beginning of the intervals. Using statistics over 22 samples of computations for $N$=1000 periods, we obtain $\lambda_1 = 0.1532 \pm 0.0002$, $\lambda_2 = -0.1930 \pm 0.0020$, where the indicated error bar is the mean-squared deflection. Lyapunov exponents for the map **T** are defined as $\Lambda_{1,2} = 2\pi\lambda_{1,2}$, and are, respectively, $\Lambda_1 = 0.9625$, $\Lambda_2 = -1.213$. Note that the largest exponent agrees well with the estimate $\Lambda \cong \ln[(3+\sqrt{5})/2] = 0.9624$, which follows from description of the dynamics on the attractor by the one-dimensional map (Section 4). Sum of the Lyapunov exponents is negative. This expresses the fact that in the course of the approach of a cloud of representative points to the attractor the volume of this cloud decreases exponentially and tends to zero. Estimate of the attractor dimension with the Kaplan – Yorke formula yields $d_L = 1 + \lambda_1 / |\lambda_2| \approx 1.793$. For the attractor as an object in the extended three-dimensional phase space, the dimension is larger by one.

## *4. Symbolic dynamics and periodic orbits*

Let us turn to the approach, known as *symbolic dynamics*. For this, the area in the Poincaré cross-section visited by the attractor trajectories, has to be subdivided in a certain way in sub-areas. Each orbit is encoded by a sequence of characters assigned to these sub-areas. It is proven that for a hyperbolic attractor the decomposition may be chosen properly to have a one-to-one correspondence between trajectories on the attractor and a set of infinite symbolic sequences composed of the characters following some "grammatical rules" (the so-called Markov partition).

For attractor of the Hunt model an appropriate partition of the domain U is shown in Fig. 7. Recall that U is a union of three domains $U_1$, $U_2$, and $U_3$. A sub-domain $U_1$ contains three partition elements, *p, a, q*. They are defined in such way that images of points from *p* and *q* under the map **T(x)** arrive again to $U_1$, while images of points from *a* to $U_3$. The borders follow coordinate lines of the curvilinear coordinate system in $U_1$ (see definition (A.1)), namely, by the equations $\theta^{(1)} = \pm(X_1 + \pi)(1 - 2/\lambda)$. Sub-domain $U_2$ consists of three elements, *x, c, y*. Images of points from *x* and *y* under the map **T(x)** arrive to $U_2$, and from *c* to $U_1$. The borders are defined by the equations $\theta^{(2)} = \pm(X_2 + \pi)(1 - 2/\lambda)$ and correspond to coordinate lines in $U_2$. Finally, the sub-domain $U_3$ consists of a single element *b*. Symbolic sequences encoding trajectories on the attractor are composed of seven symbols *p, a, q, x, c, y, b* with rules formulated below.

A useful representation of the dynamics is based on agreement not to distinguish points in the same area $U_i$, which have identical $\theta^{(i)}$. The Hant model is arranged in such way that in the Poincaré cross-section their images will remain on a common coordinate line $\theta$ for all subsequent iterations. In such interpretation, the phase space becomes one-dimensional: it can be thought of as a rubber string with three loops stretched to three nails, see Fig. 7b. (It is called *a branched one-dimensional manifold.*) One iteration of the Poincaré map corresponds to an expansion of this string and placing it in a certain way stretched to the same three nails.

Graphically, the dynamics may be illustrated with the diagram of Fig. 7c. At each coordinate axis three segments are selected; in each the coordinate $\theta^{(i)}$ for *i*=1,2,3 is plotted. (The graph looks discontinuous, but this is a defect of the manner of representation: on the branched manifold the function is actually continuous.) The iterations are represented geometrically with a Lamerey diagram. Visit of one or other piece corresponds to a certain symbol in a code of the given trajectory. Note that all branches of the plot have equal slope, in absolute value given by the constant $\lambda = (3 + \sqrt{5})/2$. From this, we conclude that the Poincaré map should have a Lyapunov exponent equal to $\ln[(3 + \sqrt{5})/2] = 0.9624$.[1].

From the diagram of Fig. 7c one can see that the "grammar" of the symbolic representation of orbits is expressed by the following rules:

$$p \to p,a,q,\ a \to b,\ q \to p,a,q,\ x \to x,c,y,\ c \to p,a,q,\ y \to x,c,y, b \to x,c,y, \qquad (2)$$

where for each symbol we indicate the characters allowable after it.

Periodic symbolic sequences correspond to periodic orbits, or cycles on the attractor. Using the rules (2), one can list all possible cycles of certain period. In particular, there are three cycles of period 1 (codes *p, q, y*), two cycles of period 2 (codes *pq, xy*)), five cycles of period 3 (codes *ppq, pqq, abc, xxy, xyy*). There are ten cycles of period 4 (codes *pppq, ppqq, pqqq, pabc, qabc, xxxy, xxyy, xyyy, xcab, ycab*), twenty-four cycles of period 5, fifty cycles of period 6, etc.

As the attractor is strange hyperbolic, all these cycles are unstable, of saddle type. Thus, to find out them in the computations, it is necessary to use some special procedures of the search. One possible approach is to iterate the one-dimensional map of Fig. 7c back in time selecting a proper branch at each step in accordance with the prescribed symbolic sequency. Then, we complement the obtained value $\theta$ with an arbitrary *r* and integrate the equations (1) in direct time for one period of the desirable cycle with these initial conditions. The result yields approximate coordinates of the point on the specified cycle that can be improved finally with the Newton method.

---

[1] Of course this is just one, the greatest exponent. The second is excluded from consideration as we pass to description of the dynamics on the one-dimensional branched manifold.

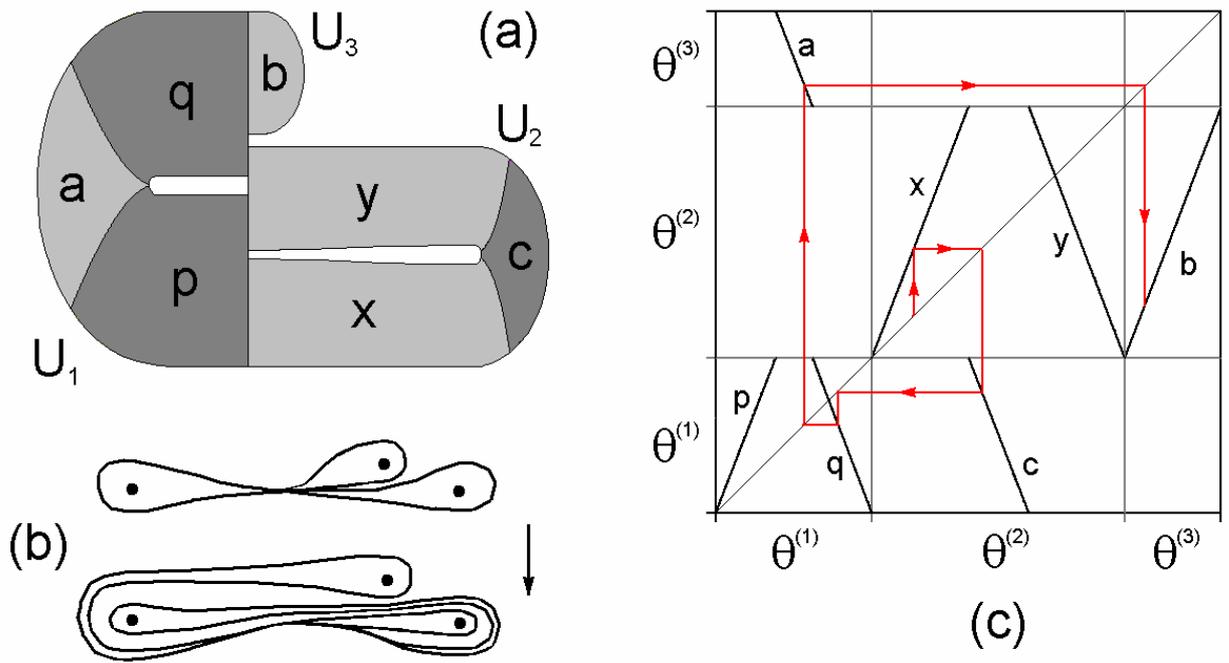

**Fig. 7.** Markov partition of the domain, in which the attractor is placed in the Poincaré cross-section (a). Representation of the dynamics with a use of the branched one-dimensional manifold (b). Representation of the dynamics on the branched manifold by means of the iteration diagram (c). A part of orbit is shown with a code *xcqab*...

**Table**

Symbolic codes, multiplies, and Lyapunov exponents for several periodic orbits of the Hunt model of period from 2 to 8

| Period | Code | $\mu_1$ | $\mu_2$ | $\Lambda_1$ | $\Lambda_2$ |
|---|---|---|---|---|---|
| 2 | *pq* | -6.857 | -0.1456 | 0.9626 | -0.9634 |
| 3 | *pqq* | 17.928 | 0.04953 | 0.9621 | -1.0017 |
| 4 | *pabc* | 46.977 | 0.004880 | 0.9624 | -1.3307 |
| 4 | *ppqq* | 47.061 | 0.004976 | 0.9628 | -1.3258 |
| 5 | *pqqpp* | 123.17 | 0.0004596 | 0.9627 | -1.5370 |
| 5 | *pabcp* | 123.11 | 0.002027 | 0.9626 | -1.2403 |
| 5 | *pabxc* | 122.89 | 0.003478 | 0.9623 | -1.1323 |
| 6 | *ppqqqq* | 322.27 | 0.0008787 | 0.9626 | -1.1728 |
| 6 | *pabcqq* | 321.90 | 0.0005048 | 0.9624 | -1.2652 |
| 6 | *pabxcp* | 322.09 | 0.001021 | 0.9625 | -1.1478 |
| 7 | *ppqqabc* | 844.70 | 0.0001869 | 0.9627 | -1.2264 |
| 7 | *ppabxcp* | 845.55 | 0.0001006 | 0.9629 | -1.3149 |
| 8 | *ppqqabcp* | 2203.1 | 0.00001750 | 0.9622 | -1.3695 |
| 8 | *pabxcpqq* | 2201.8 | 0.0001814 | 0.9621 | -1.0768 |
| 9 | *ppqqabyyc* | 5777.9 | 0.00002745 | 0.9624 | -1.1670 |
| 9 | *pabcpqppq* | 5763.7 | 0.00002374 | 0.9621 | -1.1831 |
| 9 | *ppabxyycp* | 5802.6 | 0.00001318 | 0.9628 | -1.2486 |

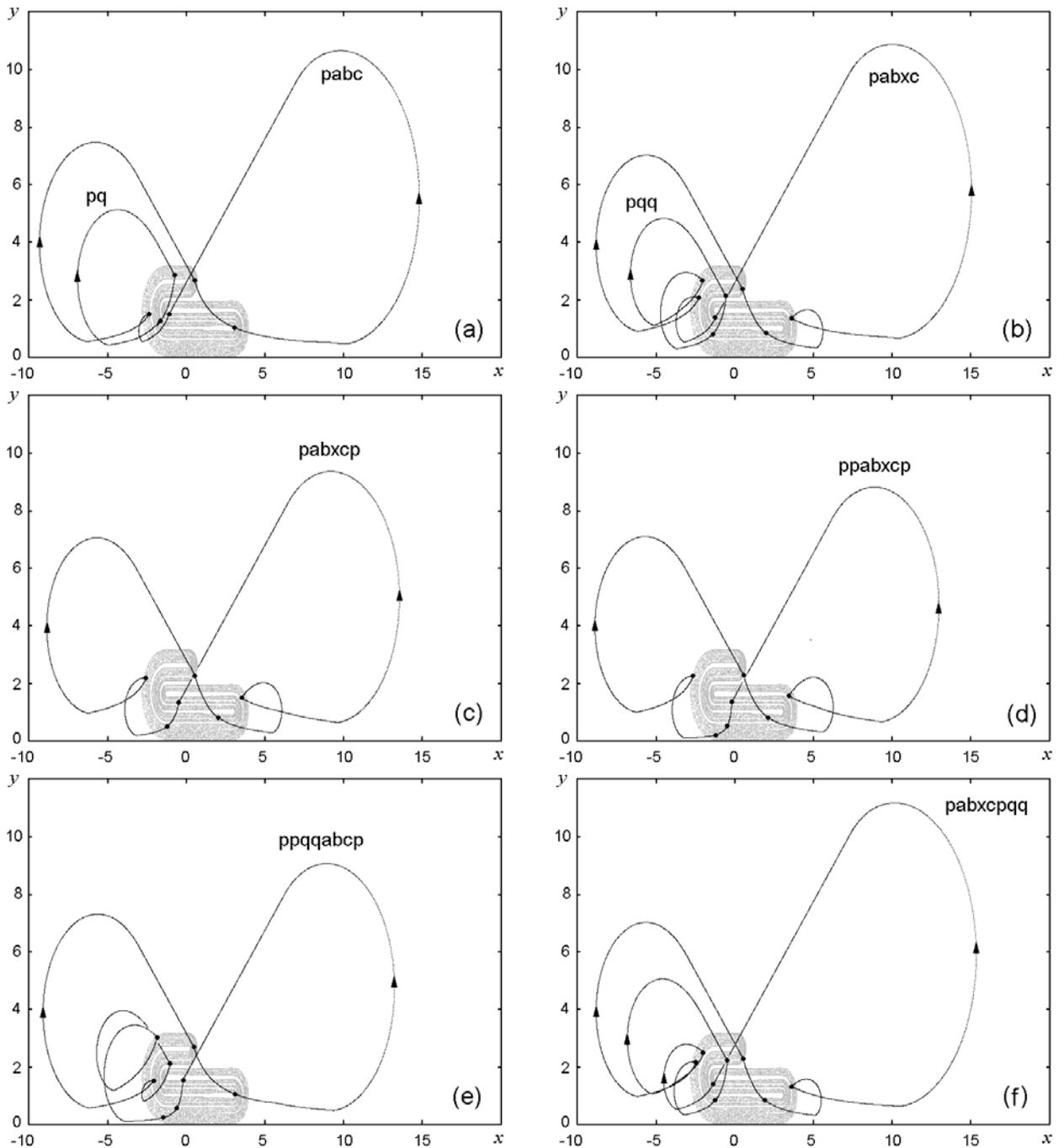

**Fig. 8.** Phase portraits of saddle cycles for the Hunt model in projection on the plane ($x,y$). The arrows indicate directions of the motion on the trajectories. Small black filled circles mark the points relating to instants $t = 2\pi n$, where $n$ is an integer, i.e. to the Poincaré cross-section. Attractor in the cross-section is shown in gray. For each periodic orbit a respective symbolic code is indicated containing a number of characters equal to the period of the orbit in units of $2\pi$.

Figure. 8 shows phase portraits for some cycles of the Hunt model of periods from 2 to 8 in projection onto the plane ($x,y$). For each cycle, the symbolic sequence is indicated. It can be restored by carefully considering the picture: black filled circles, on a background of the attractor shown in gray, are located just in the areas identified in Figure 7.

For these and some other periodic orbits, in the table we present the computed multipliers, the eigenvalues of matrices, describing transformation of a perturbation vector over a period of the orbit in the linear approximation. For each cycle one of the multipliers is always greater than one in absolute value, and the other is less than one. So, these cycles are really of the saddle type, as it must be for all orbits on the hyperbolic strange attractor. The table contains as well the Lyapunov exponents calculated for these cycles. The larger one is of almost the same value for all the cycles; up to an error of computations they coincide with $\Lambda = \ln[(3+\sqrt{5})/2] = 0.9624$. The second exponent for the different cycles varies in some limits, but it is always negative; its absolute value exceeds $\Lambda$.

## *5. Conclusion*

In this work, we reproduced construction of an artificial continuous time model system of Hunt possessing a hyperbolic strange attractor. We performed quite a detailed numerical study of the dynamics of the system, and presented phase portraits, time dependence of variables, Lyapunov exponents and dimension estimate, the Fourier power spectrum. Also we presented results of numerical verification of the cone criterion guaranteeing hyperbolic nature of the attractor. Due to this, we noted an inaccuracy in the work of Hunt and corrected it in our computations. The results of the present paper may be used in the learning process at the level of graduate and post-graduate students in radio-physics and nonlinear dynamics, as an example of a hyperbolic strange attractor to demonstrate concepts of the respective general mathematical theory. This study has contributed to accumulation of experience with hyperbolic attractors in our research group; it becomes important because of recent appearance of physically realizable systems with attractors of such type. One of interesting questions for future studies is to search for design of real systems with attractors of Plykin type, since the Hunt model itself is unlikely to allow practical implementation.

*The work was supported, in part, by the Ministry of Education and Science of Russian Federation in a frame of program of Development of Scientific Potential of Higher Education (2009-2010) in a part of Basic Research in Natural Sciences, grant No 2.1.1/1738. S.P.K. acknowledges support from RFBR-DFG grant No 08-02-91963.*

## *Appendix A*

### **Formal description of the Hunt model**

Dynamics of variables $x$, $y$ on the time interval $\Delta t = 2\pi$ is represented in three successive stages, the duration of each is $2\pi/3$. In formulation of the relations, fictitious time $s$ is used, which varies in each stage from 0 to 1. To ensure smoothness of the flow, relationship between $s$ and $t$ is chosen to have zero velocity of motion of the representative points on the plane of $x$ and $y$ at the edges of the stages.

In description of the flow, three special curvilinear coordinate systems $(r,\theta)^i$, $i=1,2,3$ are used, been linked with the Cartesian coordinates $x$, $y$ by relations

$$\begin{aligned} x &= X + S_i(\pi/2 - |\theta|), \quad y = Y + r\,\mathrm{sgn}\,\theta, \quad |\theta| \geq \pi/2, \\ x &= X + S_i r \cos f_r(\theta), \quad y = Y + r \sin f_r(\theta), \quad |\theta| < \pi/2, \end{aligned} \quad (A.1)$$

where $f_r(\theta) = 2\pi^2 r\theta/[\pi^2(r+1)+4(r-1)\theta^2]$, $S_1 = -1$ and $S_{2,3} = 1$. Variables $X$ and $Y$ determine location of the origin for the $i$-th coordinate system in the Cartesian coordinates. For brevity, we designate the transformations to new coordinates and back as $(r,\theta)^i = (R(x,y,X,Y,S_i), \Theta(x,y,X,Y,S_i))$ and $(x,y) = (F^i(r,\theta,X,Y,S_i), G^i(r,\theta,X,Y,S_i))$, respectively. The explicit expressions are derived easily from (A.1).

Let us set $\lambda = (3+\sqrt{5})/2$ and $\mu = (3-\sqrt{5})/2$, and introduce some other constants determining geometrical disposition of the construction on the plane $(x,y)$:

$$X_1 = \tfrac{1}{4}\left(6\sqrt{5} - 6 - (3-\sqrt{5})\pi\right), \quad X_2 = 3, \quad X_3 = \tfrac{1}{4}\left(18 - 6\sqrt{5} - (\sqrt{5}-1)\pi\right),$$

$$Y_1 = \tfrac{1}{2}(1+\sqrt{5}), \; Y_2 = 1, \; Y_3 = \tfrac{1}{2}(3+\sqrt{5}), \; R_1 = \tfrac{1}{2}(1+\sqrt{5}), \; R_2 = 1, \; R_3 = \tfrac{1}{2}(\sqrt{5}-1). \quad (A.2)$$

**1. On the first stage** $t \in [0, 2\pi/3]$ we set $s = \sin^2(\frac{3}{4}t)$. The origins for the special coordinates move in dependence on $s$ in accordance with

$$X_{i1}(s) = \lambda^s(X_i + \pi/2) - \pi/2, \ Y_{i1} = \mu^s Y_i, \ i = 1,2,3. \tag{A.3}$$

Besides, we set $R_{i1}(s) = \mu^s R_i$. Been given certain $s$, $x$, $y$, we determine the vector $\mathbf{f}(x,y,s) = (f(x,y,s),\ g(x,y,s))$ via the following procedure.

**(a)** If $x \leq 0$, perform transformation to the coordinate system number 1:

$$(r,\theta)^1 = \big(R(x, y, -X_{11}(s), Y_{11}(s), -1),\ \Theta(x, y, -X_{11}(s), Y_{11}(s), -1)\big), \tag{A.4}$$

and, designating by dot the derivative in respect to $s$, set

$$\dot{\theta} = (\ln\lambda)\theta, \ \dot{r} = \big(\gamma(\theta, X_{11}(s))\big)h_{12}(r, R_{11}(s)) + (1 - \gamma(\theta, X_{11}(s)))(\ln\mu)r. \tag{A.5}$$

Here the functions are introduced:

$$\gamma(\theta, X) = \begin{cases} 1, & |\theta| \leq \frac{\pi}{2}, \\ \cos^2\frac{1}{4}\pi(2|\theta| - \pi)X^{-1}, & \frac{\pi}{2} \leq \theta \leq \frac{\pi}{2} + X, \\ 0, & |\theta| \geq \frac{\pi}{2} + X, \end{cases} \quad h_{12}(r, R) = \begin{cases} \cos\pi(1 - r/\varepsilon), & r < \varepsilon, \\ 1, & \varepsilon \leq r \leq R - \varepsilon, \\ \frac{5}{4} + \frac{1}{4}\cos\frac{1}{2\varepsilon}\pi(r - R), & |r - R| < \varepsilon, \\ \frac{3}{2}, & r \geq R + \varepsilon. \end{cases} \tag{A.6}$$

Then, compute components of the vector field in Cartesian coordinates $(f_1, g_1)$ as

$$f_i = \begin{cases} \dot{X} - S\,\mathrm{sgn}(\theta)\dot{\theta}, & |\theta| \geq \pi/2, \\ \dot{X} + S\dot{r}\cos f_r(\theta) - Sr\sin f_r(\theta)\big(\dot{r}\,\partial f_r(\theta)/\partial r + \dot{\theta}\,\partial f_r(\theta)/\partial\theta\big), & |\theta| \leq \pi/2, \end{cases}$$

$$g_i = \begin{cases} \dot{Y} + \mathrm{sgn}(\theta)\dot{r}, & |\theta| \geq \pi/2, \\ \dot{Y} + \dot{r}\sin f_r(\theta) + r\cos f_r(\theta)\big(\dot{r}\,\partial f_r(\theta)/\partial r + \dot{\theta}\,\partial f_r(\theta)/\partial\theta\big), & |\theta| \leq \pi/2, \end{cases} \tag{A.7}$$

where $i=1$, $S = S_1 = -1$, $\dot{X} = -(\ln\lambda)(X_{11} + \pi/2)$, $\dot{Y} = (\ln\mu)Y_{11}$.

**(b)** If $x > 0$, perform analogous computations in the coordinate systems 2 and 3.

First, find out

$$(r,\theta)^2 = \big(R(x, y, X_{21}(s), Y_{21}(s), 1),\ \Theta(x, y, X_{21}(s), Y_{21}(s), 1)\big). \tag{A.8}$$

Then, set

$$\dot{\theta} = (\ln\lambda)\theta, \ \dot{r} = \big(\gamma(\theta, X_{21}(s) - X_{31}(s))\big)h_{12}(r, R_{21}(s)) + (1 - \gamma(\theta, X_{21}(s) - X_{31}(s)))(\ln\mu)r \tag{A.9}$$

and with the formula (A.7), where $i=2$, $S = S_2 = 1$, $\dot{X} = (\ln\lambda)(X_{21} + \pi/2)$, $\dot{Y} = (\ln\mu)Y_{21}$ compute the components of the vector field $(f_2, g_2)$.

Next, find out

$$(r,\theta)^3 = \big(R(x, y, X_{31}(s), Y_{31}(s), 1),\ \Theta(x, y, X_{31}(s), Y_{31}(s), 1)\big) \tag{A.10}$$

and set

$$\dot{\theta} = (\ln\lambda)\theta, \ \dot{r} = \big(\gamma(\theta, X_{31}(s))\big)h_3(r, R_{31}(s)) + (1 - \gamma(\theta, X_{31}(s)))(\ln\mu)r, \tag{A.11}$$

where

$$h_3(r) = \begin{cases} \cos\pi(1 - r/\varepsilon), & r < \varepsilon, \\ 1, & r \geq \varepsilon. \end{cases} \tag{A.12}$$

With the formula (A.7), where $i=3$, $S = S_3 = 1$, $\dot{X} = (\ln\lambda)(X_{31} + \pi/2)$, $\dot{Y} = (\ln\mu)Y_{31}$, obtain components of the vector field $(f_3, g_3)$.

**(c)** Now, determine the vector field

$$\widetilde{\mathbf{f}}(x,y,s) = \begin{cases} (f_1, g_1), & x \leq 0, \\ w(d_3, d_2) \cdot (f_2, g_2) + w(d_2, d_3) \cdot (f_3, g_3), & x > 0, \end{cases} \quad (A.13)$$

where the function is introduced

$$w = w(u,v) = \sin^2\left(\tfrac{1}{2}\pi u(u+v)^{-1}\right), \quad (A.14)$$

with arguments expressed as $d_\alpha = \max\{R^\alpha(x, y, X_{\alpha 1}(s), Y_{\alpha 1}(s)) - \mu^s R_\alpha, 0\}$, $\alpha=2,3$.

**(г)** As a final step of computations on the stage 1 set

$$\mathbf{f}(x,y,s) = \widetilde{\mathbf{f}}(x,y,s) + \sum_{i=1}^{3} \beta(\|\mathbf{x} - \mathbf{X}_{i1}\|) \cdot \left(\dot{\mathbf{X}}_{i1}(s) - \widetilde{\mathbf{f}}(x,y,s) + (\ln\lambda)(\mathbf{x} - \mathbf{X}_{i1})\right). \quad (A.15)$$

Here $\mathbf{x} = (x, y)$, $\mathbf{X}_{i1}(s) = (S_i X_{i1}, Y_{i1})$, $\dot{\mathbf{X}}_{i1}(s) = (S_i \dot{X}_{i1}, \dot{Y}_{i1})$, and the function is introduced

$$\beta(\rho) = \begin{cases} 1, & \rho \leq \varepsilon/4, \\ \cos^2 \pi(2\rho/\varepsilon - 1/2), & \varepsilon/4 < \rho < \varepsilon/2, \\ 0, & \rho \geq \varepsilon/2. \end{cases} \quad (A.16)$$

**Remark.** Here an inaccuracy occurs in the Hunt work. While the definitions in the text correspond to (A.15) and (A.16), in the code of the program in Mathematica the argument of the function $\beta$ is *squared* norm $\|\mathbf{x} - \mathbf{X}_{i1}\|^2$. Our computations with the procedure described in Appendix B demonstrate that this moment is significant: with substitution of the square norm the hyperbolicity disappear! On the other hand, if we follow the definitions (A.15) and (A.16), the attractor really is detected as a hyperbolic one. However, at $\varepsilon=0.05$ selected by Hunt it appears difficult to observe the fractal transversal structure of the attractor in illustrations. Thus, in our computations we take an increased parameter value $\varepsilon=0.17$, although within the allowable range specified in the work of Hunt.

**2. On the second stage** $t \in [2\pi/3, 4\pi/3]$ set $s = \sin^2(\tfrac{3}{4}t - \tfrac{1}{2}\pi)$. Taking as the origin the point

$$X_{22}(s) = X_2 + (1-s)D, \quad Y_{22}(s) = Y_2 + (1-s)D, \quad D = (\lambda-1)(X_2 + \pi/2) + \mu R_2, \quad (A.17)$$

represent an instantaneous state $(x,y)$ in new coordinates as

$$(r, \theta) = \left(R(x, y, X_{22}(s), Y_{22}(s), 1), \Theta(x, y, X_{22}(s), Y_{22}(s), 1)\right) \quad (A.18)$$

and set

$$\dot{r} = -D, \quad \dot{\theta} = D. \quad (A.19)$$

Backward transformation to Cartesian coordinates is performed with formula (A.7), where $S = 1$, $\dot{X} = -D$, $\dot{Y} = -D$, and components of the vector field $(f_0, g_0)$ are obtained. Finally, set

$$\mathbf{f}(x,y,s) = w(a_2, b_2) \cdot (f_0, g_0), \quad (A.20)$$

where the function $w$ is given by relation (A.13), and its arguments are expressed as

$$a_2 = \max\{R(x, y, X_{22}(s), \tfrac{1}{2}(Y_2 + Y_{22}(s)), 1) - \tfrac{1}{2}(-Y_2 + Y_{22}(s)) - \mu R_3, 0\},$$

$$b_2 = \max\{Y_{22}(s) - 2R_2\mu - R(x, y, X_{22}(s), Y_{22}(s), 1), 0\}. \tag{A.21}$$

**3. On the third stage** $t \in [4\pi/3, 2\pi]$ set $s = \sin^2(\tfrac{3}{4}t - \pi)$. Take the point

$$X_{13}(s) = X_1 + (1 - s)D, \quad Y_{13}(s) = Y_1 + (1 - s)D, \quad D = (\lambda - 1)(X_1 + \pi/2) + \mu R_1 \tag{A.22}$$

as the origin. From initial values $(x,y)$ perform transformation to coordinates $(r, \theta)$:

$$(r, \theta) = \left(R(x, y, -X_{13}(s), Y_{13}(s), -1), \; \Theta(x, y, -X_{13}(s), Y_{13}(s), -1)\right). \tag{A.23}$$

Next, define the flow by formula (A.19) and pass to the Cartesian coordinates by means of (A.7), setting $S = -1$, $\dot{X} = -D$, $\dot{Y} = -D$. Now,

$$\mathbf{f}(x, y, s) = w(a_3, b_3) \cdot (f_0, g_0), \tag{A.24}$$

where the function $w$ is defined by (A.14), and

$$a_3 = \max\{R(x, y, -X_{13}(s), \tfrac{1}{2}(Y_1 + Y_{13}(s)), -1) - \tfrac{1}{2}(-Y_1 + Y_{13}(s)) - \mu R_2, 0\},$$

$$b_3 = \max\{Y_{13}(s) - 2R_1\mu - R(x, y, -X_{13}(s), Y_{13}(s), -1), 0\}$$

**4. Finally**, setting $\mathbf{f}(x, y, s) = (f(x, y, s), g(x, y, s))$, we obtain the equations for the continuous-time model valid on all three stages as

$$dx/dt = \tfrac{3}{4}|\sin \tfrac{3}{2}t| f(x, y, s(t)), \quad dy/dt = \tfrac{3}{4}|\sin \tfrac{3}{2}t| g(x, y, s(t)). \tag{A.25}$$

In the right-hand parts a factor $s'(t) = \tfrac{3}{4}|\sin \tfrac{3}{2}t|$ is taken into account, arising because of the passage from fictitious time $s$ to natural time $t$.

## *Appendix B*

### Sufficient conditions of hyperbolicity and their verification

To check hyperbolic nature of the attractor, let us turn to computational verification of the *cone criterion* known from the mathematical literature [1-7, 21].

Let us have a smooth map $\bar{\mathbf{x}} = \mathbf{T}(\mathbf{x})$ that determines discrete-time dynamics on an attractor *A*. (In our case this will be the Poincaré map of the Hunt model.) The criterion requires that with appropriate selection of a constant $\gamma > 1$, for any point $\mathbf{x} \in A$, in the space of vectors of infinitesimal perturbations one can define the expanding and contracting cones $S_\mathbf{x}^\gamma$ and $C_\mathbf{x}^\gamma$. Here $S_\mathbf{x}^\gamma$ is a set of vectors satisfying the condition that their norms increase by factor $\gamma$ or more under the action of the map. $C_\mathbf{x}^\gamma$ is a set of vectors, for which the norms increase by factor $\gamma$ or more under the action of the inverse map $\tilde{\mathbf{x}} = \mathbf{T}^{-1}(\mathbf{x})$. The cones $S_\mathbf{x}^\gamma$ and $C_\mathbf{x}^\gamma$ must be invariant in the following sense. (i) For any $\mathbf{x} \in A$ the image of the expanding cone from the pre-image point $\tilde{\mathbf{x}}$ must be a subset of the expanding cone at $\mathbf{x}$. (ii) For any $\mathbf{x} \in A$ the pre-image of the contracting cone from the image point $\bar{\mathbf{x}}$ must be a subset of the contracting cone at $\mathbf{x}$.

Definition of the expanding and contracting cones depends, in general, on the coordinate system. Their violation may be linked not only with absence of the hyperbolicity, but with inappropriate selection of the coordinate system. For the Hunt model the Cartesian coordinates are not good in this sense, while a use of the curvilinear coordinates $(\theta, r)^i$ introduced in Appendix A appears successful. Note that in each of the sub-areas $U_i$, $i=1,2,3$ these coordinates are defined differently.

The verification of the conditions **for the expanding cones** consists of the following. First, been given a point $\mathbf{x} = \mathbf{x}_0 = (x_0, y_0) \in U_i$, we determine the curvilinear coordinates $(\theta_0, r_0)^i$. Next, we take the points $\mathbf{x}_1 = (\theta_0 + h, r_0)^i$ and $\mathbf{x}_2 = (\theta_0, r_0 + h)^i$, where $h \ll 1$, and compute their Cartesian coordinates $(x_1, y_1)$ и $(x_2, y_2)$ by means of (A.1).

Let us perform numerical solution of Eqs. (4) on an interval of $t$ from 0 to $2\pi$ with initial conditions $(x_k, y_k)$, $k$=0, 1, 2. The resulting state vectors will be $\mathbf{x}' = \mathbf{x}'_0$, $\mathbf{x}'_1$, $\mathbf{x}'_2$, and let $\mathbf{x}'_0 \in U^j$. By transformation to the curvilinear coordinates we get $(\theta_k, r_k)^j$, $k$=0, 1, 2, calculate the perturbation components normalized by $h$, and compose a matrix

$$\mathbf{U} = \begin{pmatrix} u_{11} & u_{12} \\ u_{21} & u_{22} \end{pmatrix} = \begin{pmatrix} (\theta'_1 - \theta'_0)h^{-1} & (\theta'_2 - \theta'_0)h^{-1} \\ (r'_1 - r'_0)h^{-1} & (r'_2 - r'_0)h^{-1} \end{pmatrix} \approx \mathbf{DT}_\mathbf{x}. \tag{B.1}$$

In similar way, starting at $\mathbf{x}' = \mathbf{x}'_0$, we obtain the matrix $\mathbf{U}' \approx \mathbf{DT}_{\mathbf{x}'}$.

A condition that vector $\mathbf{u} = (\xi, \eta)$ relates to the cone $\mathbf{T}(S^\gamma_\mathbf{x})$ may be represented as an inequality $\|\mathbf{u}\| \geq \gamma \|\mathbf{U}^{-1}\mathbf{u}\|$ or

$$\xi^2 + \eta^2 \geq \gamma^2[(\bar{u}_{11}\xi + \bar{u}_{12}\eta)^2 + (\bar{u}_{21}\xi + \bar{u}_{22}\eta)^2], \tag{B.2}$$

where $\bar{u}_{ij}$ are elements of the matrix $\mathbf{U}^{-1} = \begin{pmatrix} \bar{u}_{11} & \bar{u}_{12} \\ \bar{u}_{21} & \bar{u}_{22} \end{pmatrix} = \begin{pmatrix} u_{22}D^{-1} & -u_{12}D^{-1} \\ -u_{21}D^{-1} & u_{11}D^{-1} \end{pmatrix}$, $D = u_{11}u_{22} - u_{12}u_{21}$.

It may be rewritten as

$$a\xi^2 + 2b\xi\eta + c\eta^2 \leq 0, \tag{B.3}$$

where $a = \bar{u}_{11}^2 + \bar{u}_{21}^2 - \gamma^{-2}$, $b = \bar{u}_{11}\bar{u}_{12} + \bar{u}_{21}\bar{u}_{22}$, $c = \bar{u}_{12}^2 + \bar{u}_{22}^2 - \gamma^{-2}$.

A condition that the same vector relates to the cone $S^\gamma_{\mathbf{T}(\mathbf{x})}$ is $\|\mathbf{U}'\mathbf{u}\| > \gamma \|\mathbf{u}\|$, or

$$(u'_{11}\xi + u'_{12}\eta)^2 + (u'_{21}\xi + u'_{22}\eta)^2 \geq \gamma^2(\xi^2 + \eta^2), \tag{B.4}$$

or

$$a'\xi^2 + 2b'\xi\eta + c'\eta^2 \geq 0, \tag{B.5}$$

where $a' = u'^2_{11} + u'^2_{21} - \gamma^2$, $b' = u'_{11}u'_{12} + u'_{21}u'_{22}$, $c' = u'^2_{12} + u'^2_{22} - \gamma^2$.

As checked in the computations, at proper selection of $\gamma$ (in some bounded interval ($\gamma_1$, $\gamma_2$), where $\gamma_1<1$ and $\gamma_2>1$), for all points on the attractor, the inequalities are valid $a < 0, c > 0$ and $a' > 0, c' < 0$, as well as $b^2 - ac > 0$ and $b'^2 - a'c' > 0$. Then, the relation (B3) is true, if

$$k^{(1)} \leq \eta/\xi \leq k^{(2)}, \quad k^{(1,2)} = -b/c \pm \sqrt{(b/c)^2 - a/c},$$

that determines the cone $\mathbf{DT}(S^\gamma_\mathbf{x})$ on the plane $(\xi, \eta)$. On the other hand, the inequality (B.5) is true, if

$$l^{(1)} \leq \eta/\xi \leq l^{(2)}, \quad l^{(1,2)} = -b'/c' \pm \sqrt{(b'/c')^2 - a'/c'}$$

that determines the cone $S^\gamma_{\mathbf{T}(\mathbf{x})}$. The inclusion $\mathbf{DT}_\mathbf{x}(S^\gamma_\mathbf{x}) \subset S^\gamma_{\mathbf{T}(\mathbf{x})}$ is guaranteeing, if $l^{(1)} < k^{(1)}$ and $l^{(2)} > k^{(2)}$. A sufficient condition for this is an inequality, directly checked in the course of the computations:

$$H > Q, \tag{B.6}$$

where

$$H = \sqrt{(b'/c')^2 - a'/c'} - \sqrt{(b/c)^2 - a/c}, \quad Q = |b'/c' - b/c|. \tag{B.7}$$

It may be shown that at γ<1 the same procedure performed at points on the attractor ensures verification of the condition for **the contracting cones** with parameter $\gamma' = 1/\gamma > 1$: $\mathbf{DT}_{\mathbf{x}}^{-1}(C_{\mathbf{x}}^{1/\gamma}) \subset C_{\mathbf{T}^{-1}(\mathbf{x})}^{1/\gamma}$. It is so, because the cones $S^{\gamma}$ и $C^{1/\gamma}$ are complementary sets: $S^{\gamma} \cup C^{1/\gamma} = \mathbb{V}$.

The above computations have been performed for a set of 20000 points in the Poincaré cross-section of the Hunt model at ε=0.17.

With $\gamma^2 = 2$ it was found that $\max a \approx -0.304 < 0$, $\min c \approx 3.96 > 0$, $\min(b^2 - ac) \approx 1.898 > 0$, $\min a' \approx 3.337 > 0$, $\max c' \approx -0.311 > 0$, $\min(b'^2 - a'c') \approx 1.827 > 0$, $1.1 < H < 4.4$, $\max Q/H \approx 0.359 < 1$. With $\gamma^2 = \frac{1}{2}$ it is found that $\max a \approx -0.294 < 0$, $\min c \approx 3.85 > 0$, $\min(b^2 - ac) \approx 1.932 > 0$, $\min a' \approx 3.346 > 0$, $\max c' \approx -0.322 > 0$, $\min(b'^2 - a'c') \approx 1.731 > 0$, $1.1 < H < 4.4$, $\max Q/H \approx 0.300 < 1$.

We conclude that with γ equal $\sqrt{2}$ and $\frac{1}{\sqrt{2}}$ the inequalities required by the cone criterion are indeed valid on the attractor of the Poincaré map, and, as follows, the attractor is hyperbolic.